\documentclass{llncs}

\usepackage{times}
\usepackage{setspace}

\usepackage{graphicx}
\usepackage[small]{subfigure}
\usepackage[small]{caption}

\usepackage[english]{babel}
\usepackage{amssymb}
\usepackage{url}
\usepackage[super,negative]{nth}
\usepackage{icomma}                

\usepackage[ruled]{algorithm2e}

\renewcommand{\subsection}[1]{\vspace*{-1.5ex}\subsubsection*{#1.}}

\newcommand{\traceroute}{\url{traceroute}}
\newcommand{\as}{AS}
\newcommand{\caida}{\textsc{Caida}}
\newcommand{\skitter}{\textit{skitter}}
\newcommand{\ip}{IP}

\begin{document}

\title{Describing and Simulating Internet Routes}

\author{J\'er\'emie Leguay\inst{1}, Matthieu Latapy\inst{2}, Timur Friedman\inst{1}, Kav\'e Salamatian\inst{1}}

\institute{LIP6 -- CNRS and Universit\'e Pierre et Marie Curie\\
8, rue du Capitaine Scott, 75015 Paris, France\\
Tel. +33 1 44 27 71 34, Fax +33 1 44 27 53 53\\
\email{\{jeremie.leguay,timur.friedman,kave.salamatian\}@lip6.fr}
\and
LIAFA -- CNRS and Universit\'e Denis Diderot,\\
2, place Jussieu, 75005 Paris, France\\
Tel. +33 1 44 27 28 42, Fax +33 1 44 27 68 49\\
\email{latapy@liafa.jussieu.fr}}

\maketitle

\begin{abstract}
This paper introduces relevant statistics for the description of
routes in the internet, seen as a graph at the interface level.
Based on the observed properties, we propose and evaluate methods
for generating artificial routes suitable for simulation purposes.
The work in this paper is based upon a study of over seven million
route traces produced by \caida's \skitter\ infrastructure.
\end{abstract}

\noindent
\begin{keywordname}
Network measurements, graphs, statistical analysis, modeling, simulation.
\end{keywordname}

\section{Introduction}

Realistic modeling of routes in the internet is a challenge for
network simulation. Until now, one has had to choose one of the
three following approaches to simulate routes: (1) use the
shortest path model, (2) explicitly model the internet hierarchy,
and separately simulate inter- and intra-domain routing, or (3)
replay routes that have been recorded with a tool like
\traceroute~\cite{traceroute}.  All of these methods have serious
drawbacks.

The first method does not reflect reality: routes do not in
general have the same properties as shortest paths, as already
pointed out by
Paxson~\cite{paxson96end2endrouting,paxson97end2endrouting},
because of routing
policies~\cite{spring03quantifying,tangmunarunkit01inflation}
mainly at the autonomous system (\as) level. As described in
detail recently by Spring et al.~\cite{spring03quantifying}, and
earlier by Tangmunarunkit et
al.~\cite{tangmunarunkit01impact,tangmunarunkit01inflation}, this
often induces \emph{path inflation}.  The second method is limited
by our ability to explicitly simulate the internet hierarchy. Much
work~\cite{tangmunarunkit02generators,barford01marginal} has been
done in order to model the internet graph, and much progress has
been made, but today's topology generators are still capable of
being highly inaccurate in capturing some parameters while they
strive to adhere to others.  (See, for instance, the findings in
Li et al.'s Sigcomm 2004 paper~\cite{li04firstprinciples}.) Then,
even if one is satisfied with the quality of the topology
simulation, there is the question of simulating dynamic inter- and
intra-domain routing.  A non-negligible programming effort is
required if the choice is made not to use a simulator, such as
\emph{ns}~\cite{ns}, that has these algorithms built in.  Finally,
the third method is not suitable if routes from a large number of
sources are to be simulated.  Today's route tracing systems employ
at most a few hundred sources.  \caida's
\emph{skitter}~\cite{huffaker02topology,skitterurl}
infrastructure, for instance, produces an extensive graph suitable
for simulations, but it based on routes from just thirty sources.

Note that despite its well known drawbacks, and because of the
lack of more accurate models, the shortest path model is generally
used. Examples from recent years include Lakhina et al.'s Infocom
2003 paper~\cite{lakhina02sampling}, Barford et al.'s Sigcomm 2002
paper~\cite{barford01marginal}, Riley et al.'s \textsc{Mascots}
2000 paper~\cite{riley00stateless}, and Guillaume et al.'s Infocom
2005 paper~\cite{guillaume05relevance}.  The ns network simulator
documentation proposes simulating routes by shortest paths as an
alternative to simulating routing
algorithms~\cite[Chs.~26,~29]{ns}.

This paper's principal contribution is a new approach to modeling
routes in the internet, one that does not share the drawbacks just
described.  We suggest using an actual measured graph of the
internet topology, such as the graph generated by skitter.  From
that topology, we suggest choosing sources and destinations as one
wishes from the nodes of the graph. Between these sources and
destinations, we suggest generating artificial routes with a model
chosen to reflect statistical properties of actual routes.

Central to this contribution are two specific models for
artificial route generation: the random deviation model and the
node degree model. These models generate routes with relatively
inexpensive calculations, and the routes that they generate better
reflect the statistical properties of actual routes than does the
shortest path model.

This paper's other contribution is to update measurements of some
familiar statistical properties of real routes, notably path
length and the hop direction, and to introduce and measure a new
statistical property: the evolution of node degree along a route.
These properties serve as the standard for evaluating whether
simulated routes resemble real routes.  By introducing this
standard, this paper lays the groundwork for going beyond the work
described here through the eventual introduction of yet better
models.

The remainder of this paper is organized as follows.
Sec.~\ref{sec_framework} describes the data set that we have used
and the context in which our work lies. Sec.~\ref{sec_stats}
proposes the set of statistical properties to describe routes in
the internet. Sec.~\ref{sec_route_models} proposes the models we
use to simulate routes based on these properties.
Sec.~\ref{sec_evaluation} evaluates those models, and
Sec.~\ref{sec_conclusion} concludes the paper.

\section{The framework}\label{sec_framework}

The ideal perspective from which to characterize routes in the
internet would be from a snapshot of the routing tables of routers
throughout the network.  Unfortunately, such a snapshot is
impossible to obtain on the scale of the entire network.  In this
section, we describe the alternative that we opted for, and the
hypotheses we made.

\subsection{The internet as a graph}

Efforts to map the internet graph take place at two levels. One is
the autonomous system (\as) connectivity graph, which can be
constructed from BGP announcements (captured for instance by The
Oregon Route Views Project~\cite{routeviews}). The other is the
router and \ip\ graph, which can be obtained using traceroute and
similar tools from a number of different points in the network. To
our knowledge, skitter, which conducts traceroutes from on the
order of 30 servers to on the order of a million destinations, is
the most extensive ongoing effort at the \ip\ level.

Neither level is ideally suited to the task of modeling the
behavior of routes at the router level. While the \as\ graph is
directly based upon routing information, it is too coarse-grained
to capture the details of path inflation. For this study, we
therefore focussed on the \ip\ and router level.

The main problem with this level is that what one actually sees is
the graph of \ip\ interfaces, while the graph of routers is more
relevant.  One single node in the router graph appears as several
separate nodes, one or more for each of its interfaces, in the
\ip\ graph. Ideally, then, one would construct the router graph
using methods to ``disambiguate'' \ip\ addresses, such as the
alias resolution techniques described by Pansiot et
al.~\cite{pansiot98onroutes}, and by Govindan et
al.~\cite{govindan00heuristics} for \emph{Mercator}.  There are
also techniques, such as those used by Spring et
al.~\cite{spring02measuring,spring04aliases}, in \emph{Rocketfuel}, and by
Teixeira et al.~\cite{teixeira03insearch}, that take advantage of
router and interface naming conventions to infer router-level
topology.

We do not use the router-level graph, however.  The disambiguation
techniques, as applied for example in the \emph{iffinder} tool
from \caida~\cite{iffinder}, do not work by simple inspection of
the \ip\ graph; they require active probing, preferably
simultaneously with graph discovery.  This constraint makes
extensive disambiguated router-level graphs much harder to obtain
than \ip\ interface graphs. At best, some core network topologies
are available in this form thanks to Rocketfuel.  But Rocketfuel
is untested in stub networks. Finally, it is very difficult to
judge the extent to which disambiguation is successful, and
incomplete or incorrect disambiguation could introduce unknown
biases.

To avoid these difficulties, we have restricted ourselves to the
\ip\ graph as obtained from skitter. The resulting caveat is that
the graph may not be properly representative of the router level
graph.

This caveat is however mitigated by the fact that the \ip\ graph
nonetheless resembles the router graph in one important respect:
route lengths are preserved.  That is to say that a route that has
a given length in the router level graph has the same length in
the corresponding \ip\ graph. Furthermore, as Broido et al.\
note~\cite{connectivity}, ``interfaces are individual devices,
with their own individual processors, memory, buses, and failure
modes. It is reasonable to view them as nodes with their own
connections.''

\subsection{The data set}\label{sec_data_set}

This study uses skitter data from July \nth{2} 2003. The data was
collected from 23 servers targeting 594,262 destinations.  We
obtained the corresponding \ip\ graph by merging the results of
the 7,075,189 traceroutes conducted on that day. This graph
captures the small-world, clusterized, and scale-free nature of
the internet already pointed out for instance in numerous
publications~\cite{jin02smallworld,faloutsos99powerlaw,vasquez--internet,vasquez02largescale,adamic99small,broder00graph}.
In particular, the average distance is approximately $12.54$ hops,
and the degree distribution is well fitted by a power law of
exponent $1.97$.

Notice that this graph is necessarily incomplete and biased due in
particular to probing from a limited number of sources, to route
dynamics, to tunneling and to erroneous or absent responses to
traceroute probes. Biases of graphs induced by acquisition through
a small number of traceroute monitors have been studied for
instance in by Lakhina et al.~\cite{lakhina02sampling}.

However, recent studies by Dall'Asta et
al.~\cite{dallasta04statistical} and Guillaume et
al.~\cite{guillaume05relevance} show that one may be quite
confident of the accuracy, using this kind of exploration, of
distances and degrees, which are the main properties that we study
here. We therefore consider the \ip\ interface graph in this
study, and in particular we use the skitter data as it represents
the current state of the art in its extent and accuracy.

\section{Statistical properties of routes} \label{sec_stats}

This section presents a set of properties for statistical
description of internet routes.  These properties motivate the
models of Sec.~\ref{sec_route_models}. Several properties have
already been studied in previous work, and the work here serves to
evaluate and update them.

\subsection{Route lengths}

\begin{figure}[h!]
\centering \subfigure[Distributions of route lengths, shortest
path lengths, and their
differences]{\label{fig_length}\includegraphics[width=4.5cm]{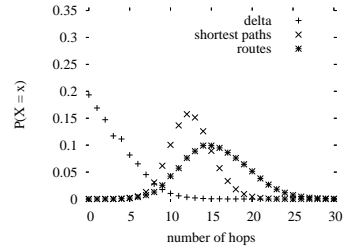}}\qquad
\subfigure[Hop direction in 15-hop routes (F: Forward, S: Stable,
B:
Backward)]{\label{fig_typeoflinks.15}\includegraphics[width=4.5cm]{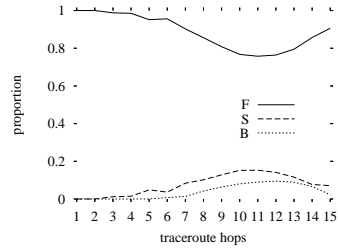}}\vspace{3pt}
\subfigure[Quantile plots for the out-degree of nodes along routes
of length
15.]{\label{fig_deg_evol}\includegraphics[width=4.5cm]{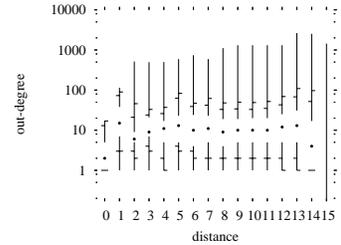}}\qquad
\subfigure[Choice of next hop as a function of its degree
ranking.]{\label{fig_deg_ch}\includegraphics[width=4.5cm]{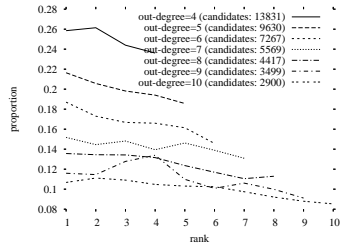}}
\caption{Statistical properties of internet
routes.}\label{fig_stats}.
\end{figure}

It is well known that routes are not shortest paths: they are not
optimal in general. Fig.~\ref{fig_length} shows the length
distributions of the routes in our data set, and of the
corresponding shortest paths. It also shows the distribution of
the difference (\textit{delta}) between the length of a route and
the corresponding shortest path. The mean length of $15.57$ hops
for routes in this data set fits closely Paxson's
observations~\cite{paxson97end2endrouting,paxson96end2endrouting}
on a data set from nine years prior.  The shortest paths have a
mean length of $12.55$ hops ($11.4$ hops if the graph is
considered to be undirected).

The delta distribution confirms Tangmunarunkit et al's
observation~\cite{tangmunarunkit01impact,tangmunarunkit01inflation},
mentioned at the beginning of this paper, that roughly 80\% of
routes are not shortest paths.  In this particular data set,
$19.34\%$ of routes are shortest paths. Moreover, since the data
is incomplete, there are undiscovered links, which implies that
$19.34\%$ is an overestimate.

\subsection{Hop direction}

When a packet travels from one router to another, it may move
closer to its destination, but also it may move farther, or it may
move to an interface that is at the same distance from the
destination as one it just left. Likewise, the distance from the
source may increase, decrease, or stay constant. We will call
these behaviors the {\em hop direction}, considered with respect
to either the destination or the source. In principle, a hop
should always increase the distance from the source and decrease
the distance to the destination; in such cases, the route is a
shortest path. Note that hop directions in the router graph can be
observed directly in the interface graph, since distances are
preserved between the two graphs.

This study observes hop direction by computing the shortest-path
hop distance from each traceroute source to all other nodes, using
breadth-first search. This is feasible due to the small number of
sources. It would also be natural to look at hop direction with
respect to the destinations but, since they are much more
numerous, it is computationally expensive.

We found that $87.3\%$ of hops go forwards, $4.6\%$ go backwards,
and $8.1\%$ remain at the same distance from the source (we call
these \emph{stable} hops).  More precisely,
Fig.~\ref{fig_typeoflinks.15} shows the portion of forward,
backward, and stable hops at each hop distance for routes of 15
hops (the most numerous ones). Note that, as one would expect, the
first and last few hops are generally forward because there are
few alternatives. On the contrary, in the core of the network a
significant proportion of the hops (more than one third) do not go
closer to the destination. This type of behavior has already been
described in the literature as the product of policy-based routing
in the core of the internet. As Tangmunarunkit et
al~\cite{tangmunarunkit01impact,tangmunarunkit01inflation} note,
such behavior may be induced by load balancing, commercial
considerations, etc.

\subsection{Degree evolution along a route}

Recent work has shown that many real-world complex networks tend
to have very heterogeneous degrees, well fitted by power laws.
This is in particular true for the internet, as observed by
Faloutsos et al.~\cite{faloutsos99powerlaw} and others. Moreover,
most of the short paths between pairs of nodes in these networks
tend to pass through the highest degree nodes. Actually, almost
all paths (not only short ones) tend to pass through these nodes,
which make them essential for network
connectivity~\cite{albert00error,kim02path,Cohen2001Attack,Cohen2000RandomBreakdown,Newman2000Robustness,guillaume04comparison}.

These observations lead us to ask how the node degree evolves
along a route.  If routes tend to pass through high degree nodes,
where do they do so, and what degree nodes do they encounter?
Furthermore, does this tendency to pass through high degree nodes
imply that, when a choice exists between next hops, the next hop
that leads to the highest degree node is generally chosen?

Fig.~\ref{fig_deg_evol} shows\footnote{In Fig.~\ref{fig_deg_evol},
dots indicate the median. Vertical lines run from the min to Q1
and from Q3 to the max. Tick marks indicate the \nth{5}, \nth{10},
\nth{90} and \nth{95} percentiles.} how node degree evolves for
routes of length 15. It reveals that a typical route does not pass
through the highest degree nodes, though a certain number of
routes do pass through some very high degree nodes. There is a
peak in median out-degree observable at distance 1. The median
falls at distance 2, rises again, and then stays fairly flat out
to distance 13, with a median degree of about 10. This leads us to
the following interpretation: the hosts have low degree, they are
connected at their first hop router to relatively high degree
nodes which play the role of access points, and then packets are
routed in a core network where the degree (typically 10) does not
depend much on the distance from the source or from the
destination.

One may wonder if there is a simple local rule that can be
observed for the degree evolution. In particular, if there is a
choice of next hop interface along a route, is there a correlation
between the degree rank of an interface and its probability of
being chosen? For instance, are higher degree interfaces chosen
preferentially over lower degree ones? Note that such a rule could
be perfectly compatible with the observed flat degree evolution.

Fig.~\ref{fig_deg_ch} plots the probability that a packet travels
to an interface's $i$-th ranked neighbor, where the neighbors are
ranked from highest out-degree to lowest.  An interface's
neighbors are its possible next hops in the directed graph. In
order to preserve the greatest detail in this middle range, the
figure does not show curves for degrees 2 or 3, or above 10, but
the curves shown are typical.

For instance, when an interface has five possible next hops, the
probability that the next hop along a route will be the highest
ranked neighbor is $0.22$.  The probability that the next hop will
be the second ranked neighbor is $0.21$.  Probabilities continue
to decrease, and the fifth ranked neighbor is chosen with a
probability of $0.18$. One can see a clear bias towards highest
degree nodes, though this bias is rather small.


\section{Route models}\label{sec_route_models}

The previous section provides a set of simple statistical tools to
capture some properties of routes in the internet. We now propose
three simple models (only two of which we eventually retain)
designed to capture these features. Each model is based upon one
statistical property studied in the previous section. Our approach
is to model a property in a very simple way and then use other
statistics to validate or invalidate the model.

Whereas our study of route properties was in the context of the
directed graphs produced by traceroute, the models in this section
are proposed for undirected graphs. The graphs available for
simulation purposes, notably those produced by topology generators
representing the router-level topology, are typically undirected
graphs. Therefore, our models must be suitable for use in this
context.

\subsection{Path length model}

The path length model is the simplest and the most obvious one
conceptually, but it proves to be unusable in practice. The model
aims at producing routes of the same lengths as real ones. As
discussed in Sec.~\ref{sec_stats}, a real route length typically
exceeds that of the shortest known path by some small integer
value $\delta \geqslant 0$.

In order to construct a route from a source $s$ to a destination
$d$, the path length model first computes the length $\ell$ of a
shortest path from $s$ to $d$. Then it samples a deviation
$\delta$ from a distribution such as the one shown in
Fig.~\ref{fig_length}, and a route is generated by choosing a path
at random from $s$ to $d$ among the ones which are loop-free and
have length $\ell+\delta$. This ensures that the difference
between shortest path lengths and actual route lengths will be
captured by the model.

To choose such a path at random implies however that one must
construct all of the loop-free paths of length $\ell + \delta$
from $s$ to $d$.  In practice, the computation required to
generate this number of paths may be prohibitive, since even in
simple cases it is exponential in $\ell + \delta$. For example, in
trying to generate all paths of length $21$ between a pair of
nodes in the skitter graph, we enumerated 1,206,525 possible
paths. Therefore, despite its simplicity, we will not consider
this model further.

\subsection{Random deviation model}

The random deviation model is based upon the idea that a route
usually follows a shortest path, but might occasionally deviate
from it. We modeled this using one single parameter, $p$, the
probability at any point of deviating from the current shortest
path to the destination, if such a deviation is possible.  We
tuned the value of $p$ to generate routes of realistic length. For
the undirected version of the skitter graph, we found $p=0.2$ to
work well.

A random deviation route from source $s$ to destination $d$ is
therefore based upon a shortest path $u$ from $s$ to $d$.  At each
hop, with probability $1-p$, the route continues along $u$. But
with probability $p$ it will, if possible, deviate off $u$ to
another path. A deviation from current node $x$ to a neighboring
node $y$ is deemed possible only if there is a shortest path $w$
from $y$ to $d$ that does not pass through $x$. Should there be a
deviation, the route continues along $w$ to $d$ (unless another
deviation should occur). The model is precisely described by
Algorithm~\ref{algo_rand_dev_route}.

Note that large numbers of routes to a given destination $d$ can
be efficiently generated with the random deviation model once a
shortest path tree rooted at $d$ has been computed.

\begin{algorithm}[!h]
\scriptsize

\linesnumbered

\SetVline

\SetKwInOut{Input}{Input}

\SetKwInOut{Output}{Output}

\SetKwInOut{Function}{Function}

\Input{A network $G$, a source $s$, a destination $d$, a deviation
probability $p$.}

\Output{An artificial route $v$ from $s$ to $d$ in $G$, following
the random deviation model.}

\Function{$\mbox{sp}(x,y)$ returns the set of all the shortest paths from $x$
to $y$ in $G$.}

\Begin{
    $u \leftarrow \mbox{random element of } \mbox{sp}(s,d)$\label{algo_rand_dev_route_initialize_u}\;
    $v \leftarrow $ empty list\label{algo_rand_dev_route_v_empty}\;
    copy the first element of $u$ to the end of $v$\label{algo_rand_dev_route_push_pop_1}\;
    remove it from $u$\;
    \While{the last element of $v$ is not $d$}{
        \If{$\mathrm{rand}[0,1] \leqslant p$\label{algo_rand_dev_route_probability_p}}{

  $C \leftarrow $ set of all the
                  shortest paths from any neighbor of $v$ to $d$\;
  Remove from $C$ the paths containing the last element of $v$\;
  \If{$C \not= \emptyset$}{
   $u \leftarrow $ random element of $C$\;
   }
        }
    copy the first element of $u$ to the end of $v$\label{algo_rand_dev_route_push_pop_2}\;
    remove it from $u$\;
    }
    return $v$\;
} \caption{$\mathrm{rand\_dev\_route} \left( G,s,d,p \right)$
\label{algo_rand_dev_route}}

\end{algorithm}

\subsection{Node degree model}

Several previous
authors~\cite{walsh99search,kim02path,bampis04short} have tried to
use the heterogeneity of node degrees to compute short paths in
complex networks. The basic idea is that a path which goes
preferentially towards high degree nodes tends to see most nodes
very rapidly (a node is considered to be seen when the path passes
through one of its neighbors).

The node degree model is based upon a similar approach, as
follows. Two paths are computed, one starting from the source and
the other from the destination. The next node on the path is
always the highest degree neighbor of the current node. The
computation terminates when we reach a situation where a node is
the highest degree neighbor of its own highest degree neighbor.
One can show that this is the only kind of loop can occur. Then,
one of two cases applies: either the two paths have met at a node,
or they have not. In the first case, the route produced by the
model is the discovered path (both paths are truncated at the meet
up node, and are merged). In the second case, we compute a
shortest path between the two loops, and then obtain the route by
merging the two paths and this shortest path, removing any loops.

This method has already been proposed~\cite{bampis04short} as an
efficient way to compute short paths in complex networks in
practice: the obtained paths are very close to shortest ones.
Moreover, the computation of the tree-like structure where each
node points to its highest degree neighbor is very simple and only
has to be processed once. Likewise, the shortest paths between a
small number of loops are computed only once. The overall model is
described in Algorithm~\ref{algo_node_degree_route}.

\begin{algorithm}[!h]
\scriptsize

\linesnumbered

\SetVline

\SetKwInOut{Input}{Input}

\SetKwInOut{Output}{Output}

\SetKwInOut{Function}{Function}

\Input{A network $G$, a source $s$, a destination $d$.}

\Output{An artificial route $v$ from $s$ to $d$ in $G$, following
the node degree route model.}

\Function{reverse$(p)$: returns the path obtained by reading $p$ from the
end to the beginning.\\
climb\_degrees$(G,v)$: returns the path in $G$ obtained from $v$
by going to the highest
degree neighbor at\\ \hspace*{2.148cm}  each hop, until it loops.}

\Begin{
    $p_s \leftarrow \mathrm{climb\_degrees}\left( G,s \right)$\;
    $p_d \leftarrow \mathrm{climb\_degrees}\left( G,d \right)$\;
    \If{$p_s$ and $p_d$ meet up}{
     let $u$ be the first node they have in common\;
     remove from $p_s$ all the nodes after $u$\;
     remove from $p_d$ all the nodes after $u$\;
     $p \leftarrow (p_s,\mathrm{reverse}(p_d))$\;
     return $p$\;
     }
    $q \leftarrow$ random shortest path from the last node of $p_s$
    to the one of $p_d$\;
    $p \leftarrow (p_s,q,\mathrm{reverse}(p_d))$\;
    remove loops from $p$\;
    return $p$\;
} \caption{$\mathrm{node\_deg\_route}\left( G,s,d \right)$}
\label{algo_node_degree_route}

\end{algorithm}

\begin{figure}[!h]
\centering \scalebox{0.7}{\input{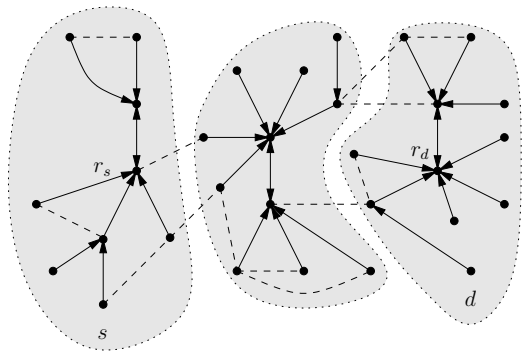}}
\caption{The node degree model: example.}
\label{fig_node_degree_model}
\end{figure}

Fig.~\ref{fig_node_degree_model} is an example. There are three
tree-like structures (the shaded areas). The source $s$ belongs to
the leftmost one, which is rooted at $r_s$, and the destination
$d$ to the rightmost one, with root at $r_d$. Each directed link
goes from one node to its highest degree neighbor (the dotted
lines are links which do not satisfy this). When one wants to
build a route from $s$ to $d$ according to the node degree model,
one first finds the path from $s$ to $r_s$, and the one from $d$
to $r_d$. One then has to compute a shortest path from $r_s$ to
$r_d$, which has length $5$ in this example. The final route is
obtained by merging these paths, and then removing the loops
(which leads to the removal of a link, in our example). It has
length $7$ (while the shortest path has length $6$).


\section{Evaluation}\label{sec_evaluation}


This section compares the performance of the random deviation and
node degree models to that of the shortest path model.  We use
undirected version of the skitter graph described in
Sec.~\ref{sec_data_set}, considered as an undirected graph. For
each model, we chose at least 60,000 (source, destination) pairs
at random from amongst the nodes of the graph and generated an
artificial route from the source to the destination. We compute
the same statistics on these routes as we had computed for actual
routes in Sec.~\ref{sec_stats}.

\begin{figure}[!htb]
\centering \subfigure[Lengths
(r.d.)]{\label{SourcesAreSkitterServers.random.lengths}\includegraphics[width=4cm]{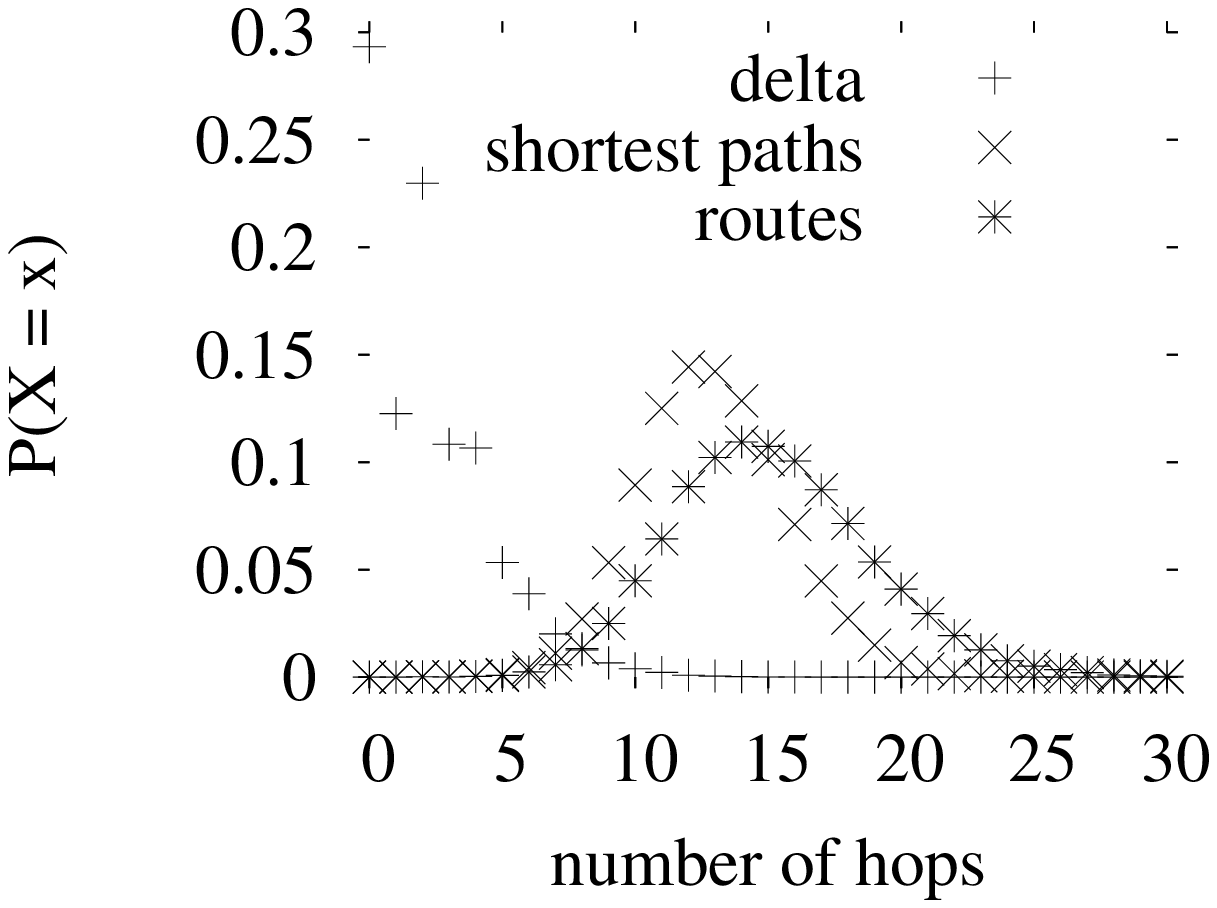}}
\subfigure[Lengths
(n.d.)]{\label{SourcesAreSkitterServers.tree.lengths}\includegraphics[width=4cm]{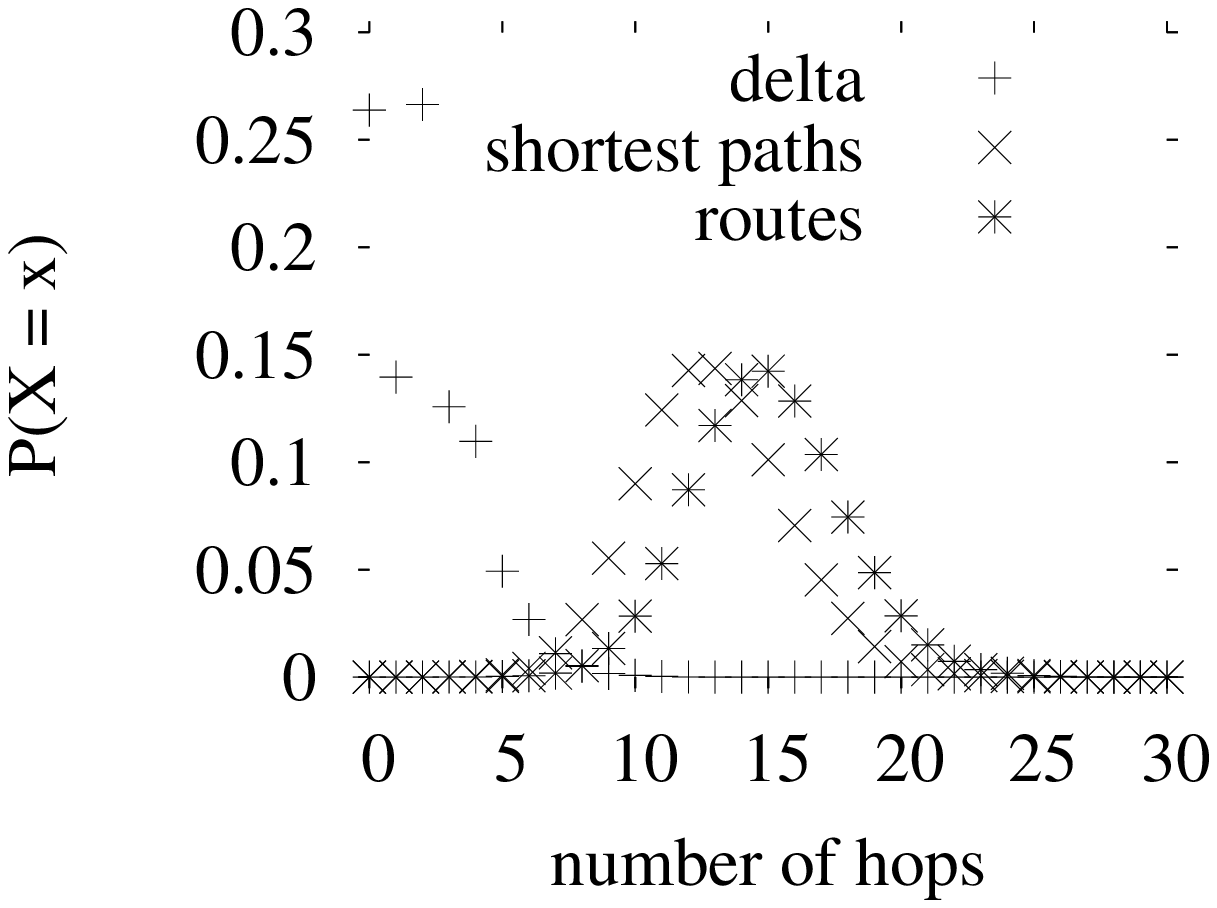}}
\subfigure[Lengths
(s.p.)]{\label{SourcesAreSkitterServers.pcc.lengths}\includegraphics[width=4cm]{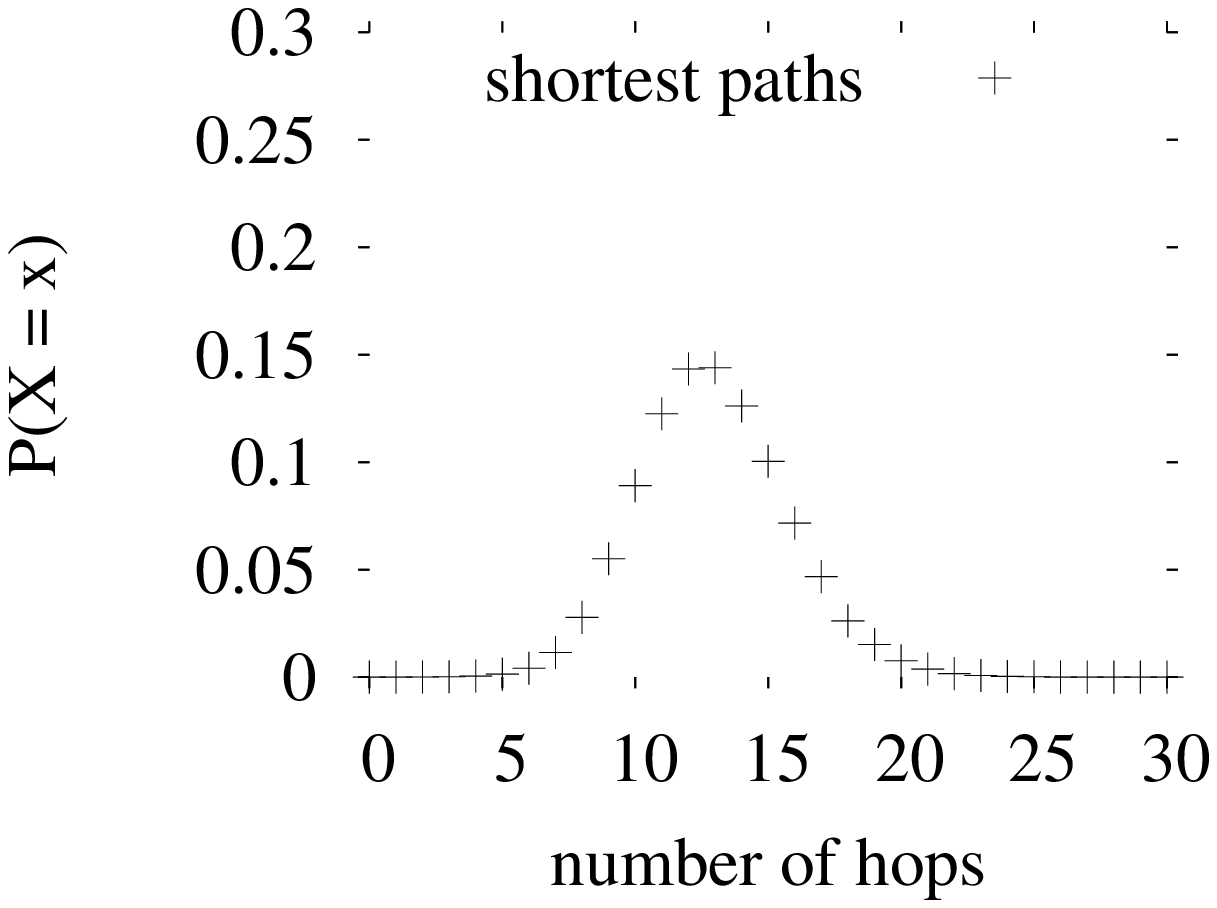}}
\subfigure[Hop direction
(r.d)]{\label{SourcesAreSkitterServers.random.16.nonunique}\includegraphics[width=4cm]{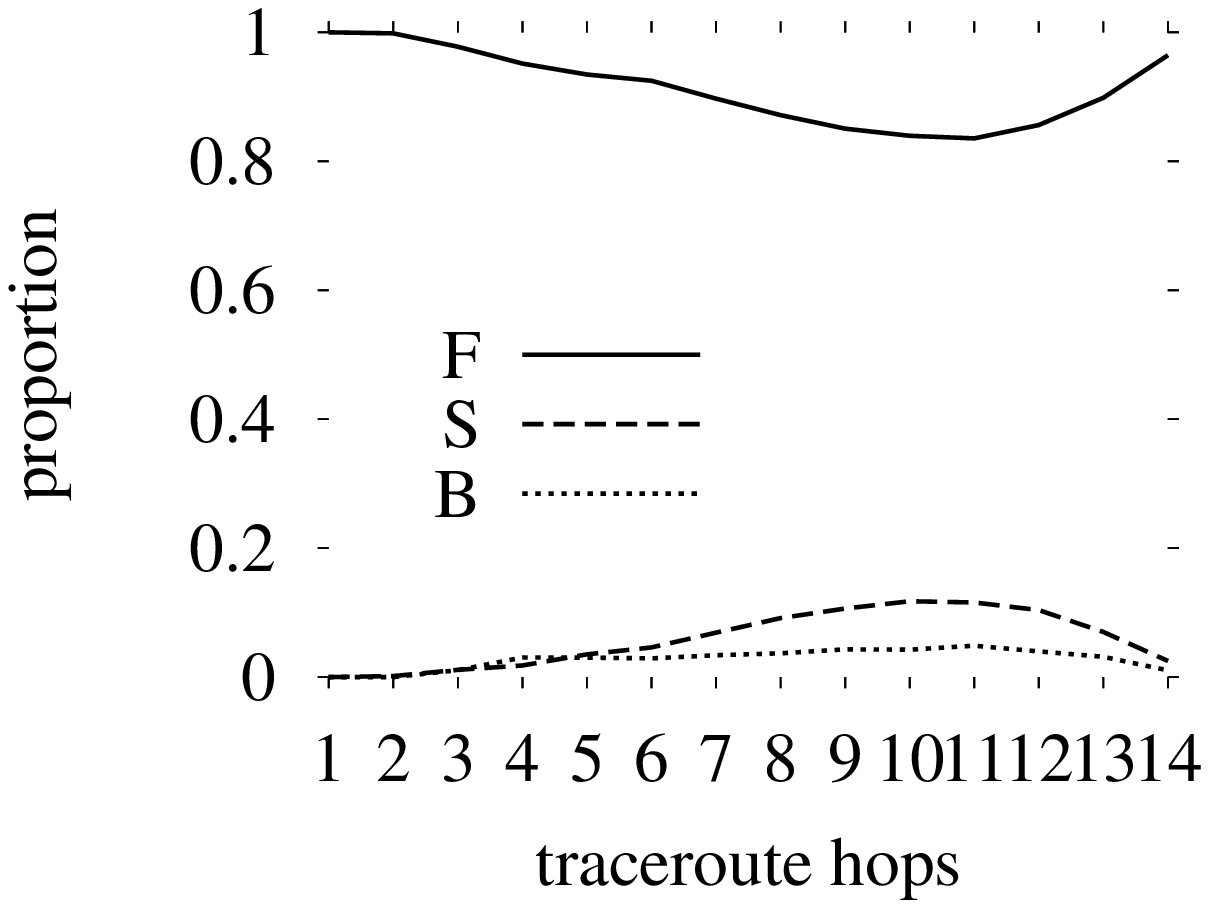}}
\subfigure[Hop direction
(n.d.)]{\label{SourcesAreSkitterServers.tree.16.nonunique}\includegraphics[width=4cm]{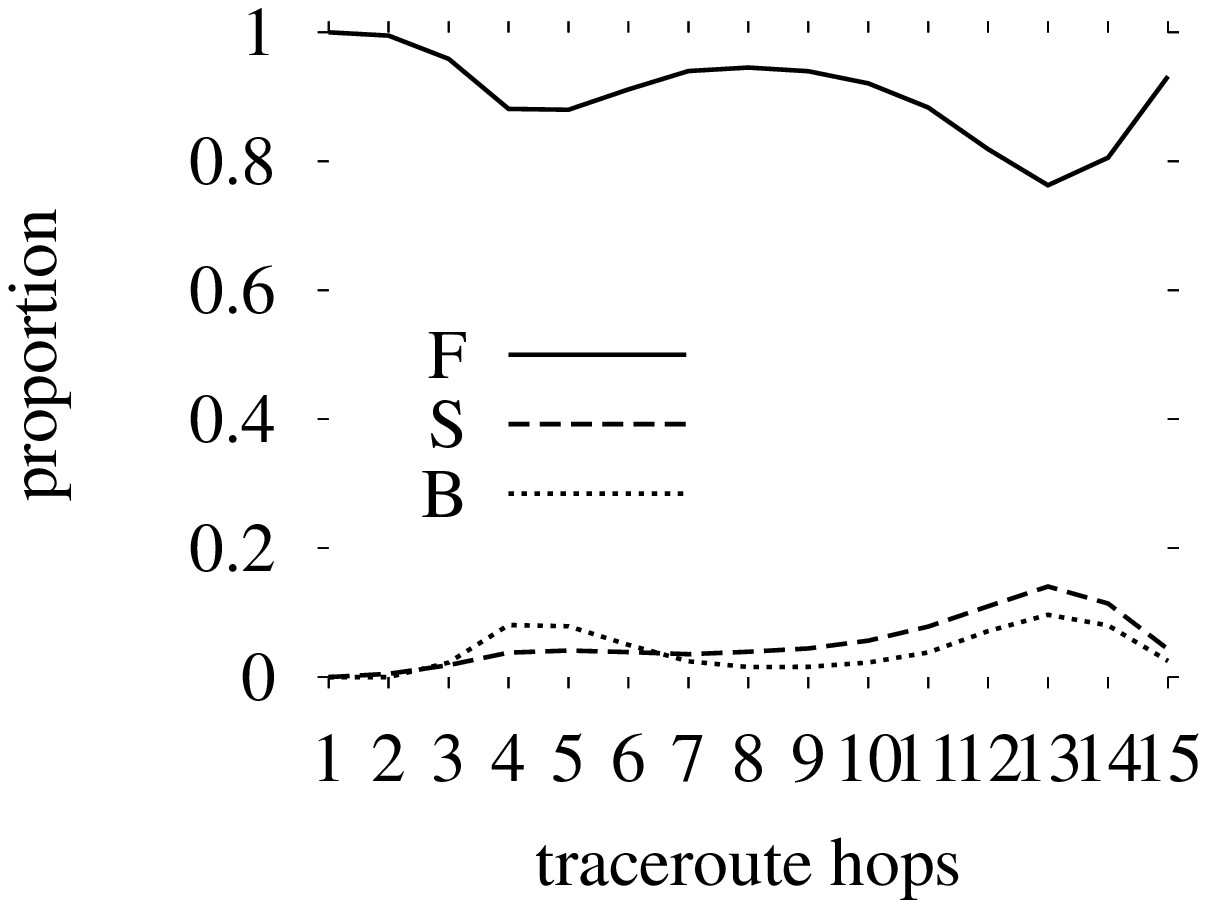}}
\subfigure[Hop direction
(s.p.)]{\label{SourcesAreSkitterServers.pcc.16.nonunique}\includegraphics[width=4cm]{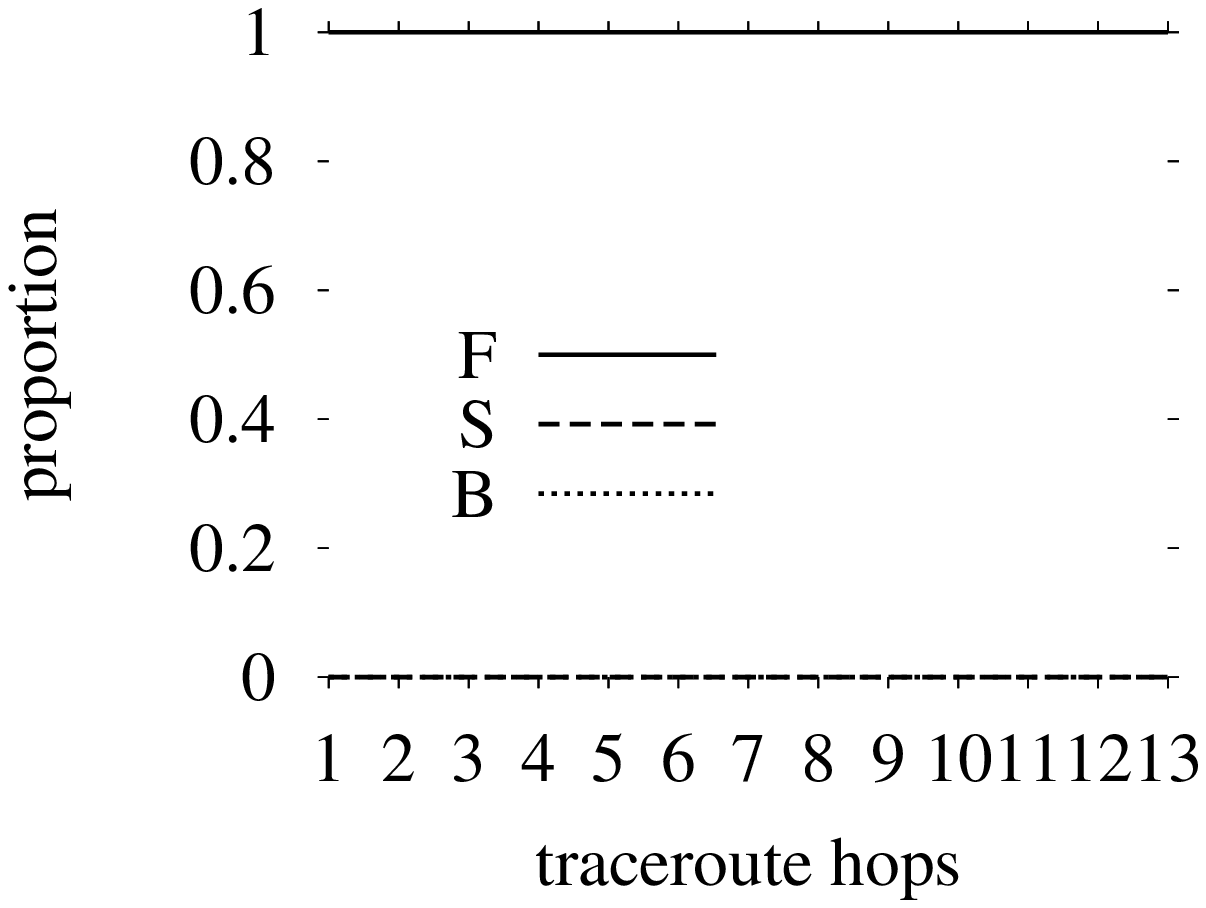}}
\subfigure[Out-degree
(r.d.)]{\label{SourcesAreSkitterServers.random.deg_evo_d}\includegraphics[width=4cm]{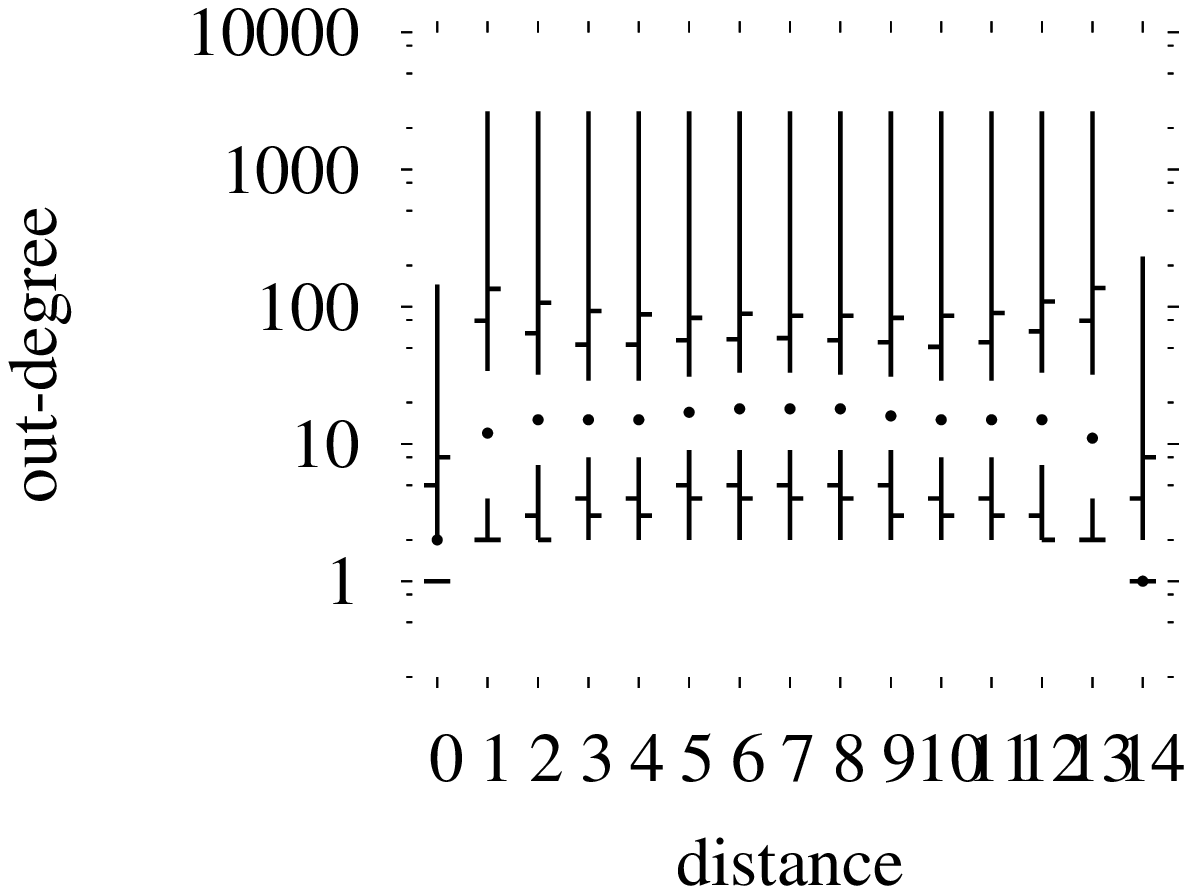}}
\subfigure[Out-degree
(n.d.)]{\label{SourcesAreSkitterServers.tree.deg_evo_d}\includegraphics[width=4cm]{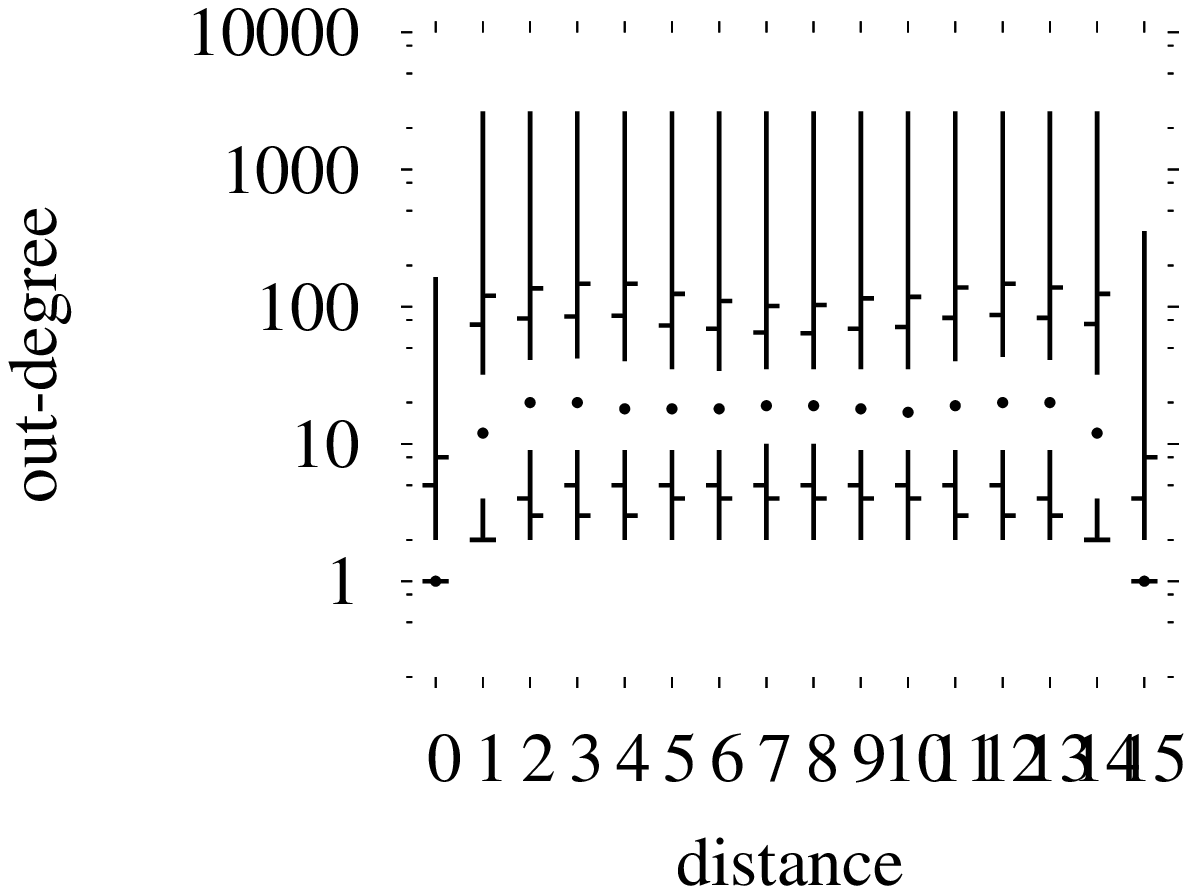}}
\subfigure[Out-degree
(s.p.)]{\label{SourcesAreSkitterServers.pcc.deg_evo_d}\includegraphics[width=4cm]{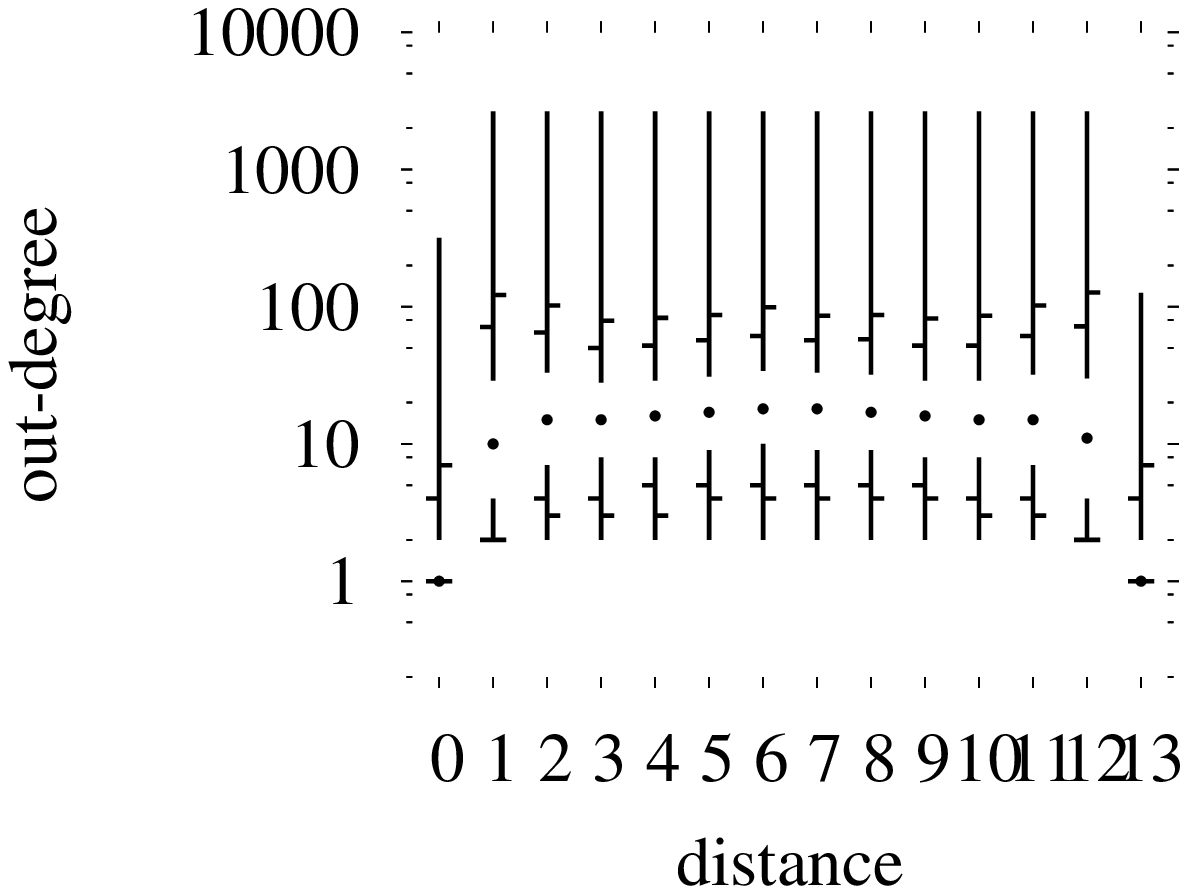}}
\caption{\label{fig_SourcesAreSkitterServers}Experiments using the
{\em random deviation model} (left), the {\em node degree model}
(center), and the {\em shortest path model} on the undirected
skitter graph using sources and destinations chosen at random from
amongst all the nodes in the graph.}
\end{figure}

Fig.~\ref{fig_SourcesAreSkitterServers} shows the statistics for
each model.  We judge the quality of a model by how well its
statistics mirror those for actual routes, shown in
Fig.~\ref{fig_stats}.

Comparing the route length distributions, we find that both models
generate distributions that are symmetric, average somewhat higher
than the shortest path distribution, and have tails similar to the
actual route length distribution shown in Fig.~\ref{fig_length}.
Mean route length is $15.15$ for the random deviation model and it
is $14.96$ for the node degree model, whereas the mean shortest
path is $12.93$.  (Note that, on the undirected skitter graph,
shortest paths between random sources and destinations are longer
on average than those between skitter sources and destinations,
for which we had computed an average route length of $11.21$.)

Lengths of paths generated with the node degree model tail off
somewhat quicker than in reality (approaching zero closer to
length 20 than length 25), but the degree of fidelity is
nonetheless remarkable given that the length distributions are not
explicitly part of the model.  The random deviation model
generates more routes that are shortest paths than in reality
(roughly 30\% compared to roughly 20\%), whereas the node degree
model generates somewhat fewer (roughly 26\%). As is already
known, the shortest path model does not capture the length
properties.

Looking at the hop directions for the most frequent route length,
we found that the curves for the random deviation model better
match the shapes of the curves for real routes shown in
Fig.~\ref{fig_typeoflinks.15}. Hops are mostly forward near the
source, but dip to around $80\%$ roughly ten hops out (whereas in
reality the portion of forward hops dips to around $80\%$ at
eleven or twelve hops out). This is in marked contrast to hop
directions produced by the node degree model because forward hops
dip much sooner and a bit less steadily. But overall portions of
forward, stable, and backward hops closely match reality for both
models: 89\% forward, 7\% stable, and 4\% backward for the random
deviation model, and 90\% forward, 6\% stable, and 4\% backward
for the node degree model, compared to 87\% forward, 8\% stable,
and 5\% backward for true routes. The shortest path model fails to
capture these proportions since all of its links are forward.

The node degree model shines compared to the random deviation
model in capturing the evolution of the out-degree close to a
route's source.  Routes generated with this model show the peak in
the out-degree before settling down to a median around 20 that we
noticed in Fig.~\ref{fig_deg_evol}, though the peak is reached at
distance 2 rather than at the first hop router. The random
deviation model and the shortest path model also have a median
around 20, but they arrive there through a smooth increase, with
no clear peak.

Based upon this comparison to real routes, we can state that the
random deviation and node degree models do a reasonable job of
emulation, though each model captures some aspects better than
others, and their strengths are different. Both models clearly
out-perform the shortest path model.


\section{Conclusion and future work}\label{sec_conclusion}


The main contribution of this paper has been to propose a new
alternative for the simulation of routes in the internet: the use
of simple models that capture non-trivial statistical properties
of routes.  The models proposed here have been found to reproduce
a number of aspects of true internet routes, though neither fully
captures all of the characteristics.  Our goal was to introduce
simple models that could serve as alternatives to the clearly
unrealistic shortest path model.  No model can be fully faithful
to reality, and the key is to understand in what ways it is a true
representation, and in what ways it diverges.  Future work along
these lines might include the development of models that
explicitly incorporate some additional characteristics, such as
the clustering coefficient. Other work might involve studying
whether certain variants on the models, such as a hybrid of the
random deviation and node degree approaches, would be more like
real routes.  Any such work must keep in mind the desirability of
keeping the models conceptually simple, easy to implement, and
computationally tractable.

We have shown how routes can be simulated on a measured graph at
the interface level. We have chosen the undirected variant of the
skitter graph, as undirected graphs are more readily available for
simulation purposes. We have also introduced the simplifying
assumption that any node can be either source or destination.
Using graphs that contain direction information and that label
end-hosts separately from routers could potentially improve the
quality of the models.  Also, these same models can potentially be
used on synthetic graphs that are meant to represent the internet
topology.

Another area into which this work could be extended would be to
capture something of the dynamics of internet routes.  There are
effectively random choices to be made in both the random deviation
model (clearly) and the node degree model (when it comes to
choosing among two or more neighbors of highest degree, or
choosing a shortest path between two trees).  Therefore these
models may produce routes that vary between a given source and a
given destination.  But we have not touched on the timing of that
variation, the topological relatedness or lack thereof of
consecutive routes, or the manner in which path lengths change
over time. Much remains to be done in this promising direction.

{\bigskip \noindent {\bf Acknowledgments.} We thank k claffy and
the staff at \caida\ for making the skitter data available to us.
We also thank Mark Crovella and Miriam Sofronia for helpful
comments. This work was supported in part by the RNRT through the
Metropolis and MetroSec projects. }

\begin{spacing}{0.05}
\bibliographystyle{plain}
\bibliography{xbib}
\end{spacing}

\end{document}